\documentstyle[preprint,aps]{revtex}
\begin{document}
\draft

\tightenlines
\title{Multishelled Gold Nanowires}

\author{G. Bilalbegovi\'c}

\address{Department of Physics, University of Rijeka,
Omladinska 14, 51 000 Rijeka, Croatia}

\date{to be published in Molecular Simulation}
 
\maketitle 

\begin{abstract}
 
The current miniaturization of electronic devices raises many questions about 
the properties of various materials at nanometre-scales. Recent molecular 
dynamics computer simulations have shown that small finite nanowires of gold 
exist as multishelled structures of lasting stability. These classical 
simulations are based on a well-tested embedded atom potential. Molecular 
dynamics simulation studies of metallic nanowires should help in developing 
methods for their fabrication, such as electron-beam litography and scanning 
tunneling microscopy.

\end{abstract}


\clearpage

\section{Introduction}

One-dimensional and quasi one-dimensional metallic structures are often used in 
various electronic devices. Wires of nanometre diameters and micrometre lengths 
are produced and studied for some time. Recent advances in experimental 
techniques, such as Scanning Tunneling Microscopy (STM) 
\cite{Ohnishi,Yanson} and 
electron-beam litography \cite{Hegger}, 
are giving rise to fabrication of wires with 
nanometre lengths. Many important results were recently obtained for nanowires 
of different materials. For example, multishelled nanostructures were found in 
experiments for carbon \cite{Iijima}, 
$WS_2$ \cite{Tenne}, $MoS_2$ \cite{Margulis}, and $NiCl_2$ \cite{Hacohen}. 
In the jellium model calculation multishelled structures were obtained 
for sodium nanowires \cite{Yannouleas}. 

The results of Molecular Dynamics (MD) simulation \cite{Goranka}
have shown that a gold 
wire with length $4$ nm and radius of $0.9$ nm, at $T=300$ K, 
consists of the three 
coaxial cylindrical shells and the thin core. Here we present an analysis of two 
additional multiwalled gold nanowires. We also propose that unusual strands of 
gold atoms recently formed in STM and observed by a transmission electron 
microscope (TEM) \cite{Ohnishi} 
are the image of cylindrical walls of multishelled gold 
nanowires. An explanation of the thinning process for the STM supported 
multishelled gold nanowires is given.

\section{Method}

To simulate metals by the classical MD method one should use many-body 
potentials. Several implementations of these potentials are available, as for 
example ones developed within the embedded-atom and effective medium theories 
\cite{Daw}. 
Gold nanowires were simulated using the glue realization of the embedded 
atom potentials \cite{Furio}. 
This potential is well-tested and produces a good 
agreement with diversity of experimental results for bulk, surfaces, and 
clusters. In contrast to most other potentials, it reproduces different 
reconstructions on all low-index gold surfaces \cite{Furio}. 
Therefore, it is expected 
that simulated gold nanowires of more than $\sim 50$ atoms realistically model 
natural structures. A time step of $7.14 \times 10^{-15}$ s was employed in simulation. 
The temperature was controlled by rescaling particle velocities. 

We started from ideal face centered cubic nanowires with the (111) oriented 
cross-section at $T=0$ K, and included in the cylindrical MD boxes all particles 
whose distance from the nanowire axis was smaller than $1.2$ nm for the first 
nanowire, and $0.9$ nm for the second one. 
The initial lengths of nanowires were $6$ 
and $12$ layers, whereas the number of atoms were $689$ and $784$. 
The samples were 
first relaxed, then annealed and quenched. To prevent melting and collapse into 
a drop, instead of usual heating to $1000$ K used in MD simulation of gold 
nanostructures, our finite nanowires were heated only to $600$ K. Such a procedure 
gives the atoms a possibility to find local minima and models a constrained 
dynamical evolution present in fabricated nanowires. The structures were 
analyzed after a long MD run at $T=300$ K.

\section{Results and Discussion}

Figure 1 shows the shape of the MD box for a nanowire of $784$ atoms 
after $7.1$ ns 
of simulation at $T=300K$.

Top views of the particle trajectories in the whole MD boxes for two nanowires 
are shown in Fig. 2 and Fig. 3. While the presence of a multishelled structure 
is obvious, after $10^6$ time steps of simulation the walls are still not 
completely homogeneous. Several atoms remain about the walls. Three cylindrical 
shells exist for the nanowire shown in Fig. 2.

The nanowire presented in Fig. 3 consists of the two coaxial near walls and a 
large filled core. The filled core is well ordered and its parallel vertical 
planes are at the spacing of $0.18$ nm. This double-walled structure suggests an 
application of similar gold nanowires as cylindrical capacitors. Therefore, we 
calculated the capacitance of finite nanometre-scale cylindrical capacitors and 
found the values of the order of $0.5$ aF for the sizes for which multishelled 
nanowires appear in simulations \cite{Josip}.

As always in computer simulations of real materials, it is important to compare 
results with experiments. Gold nanostructures were the subject of recent STM 
studies \cite{Ohnishi,Yanson}. 
Unusual strands of gold atoms (down to one row) were 
simultaneously observed by an electron microscope \cite{Ohnishi}. 
The structure of the 
strands and understanding of the thinning process for these tip-supported 
nanostructures were left for future studies. The MD trajectory plots of atoms in 
the vertical slice of the box shown in Fig. 4 resemble strands of gold atoms in 
Fig. 2 of Ref. \cite{Ohnishi}. 
Therefore, we propose that strands formed in STM are the 
image of the cylindrical walls of multishelled gold nanowires \cite{Japan}. 
Figure 4 
shows that defects exist on some rows. Gold rows with a defect were sometimes 
observed by electron microscope (see Fig. 2 (a) in Ref. \cite{Ohnishi}).

The thinning process for a multishelled nanowire should start from its central 
part, either the filled thin core (Fig.1 from Ref. \cite{Goranka}), 
the central part of a 
large filled core (Fig. 3 here), or an empty interior cylinder which first 
shrinks into a thin filled core (Fig. 2 here). When this central part is removed 
by diffusion of atoms to the tip, the next interior cylindrical wall shrinks 
into a new core, and then the process repeats. Therefore, the number of rows 
decreases by one as observed in the experiment \cite{Ohnishi}. 
At a final stage of the 
shrinkage processes an empty cylinder shrinks to one row of atoms. For 
multishelled nanowires the shrinkage of an internal cylinder is followed by the 
decrease of the radii of external cylinders and the whole nanostructure thins 
down with time. A mechanism of plastic deformation cycles of filled nanowires 
was suggested to explain the shape of necks formed in STM \cite{Untiedt}. 
In this model 
plastic deformation starts from the central cylindrical slab of a filled 
nanowire which acts as a weakest spot. For multishelled nanowires a such central 
cylindrical weak spot most often naturally forms. In STM/TEM experiments 
\cite{Ohnishi} 
it was noted that the gap between the dark lines of their Fig. 2 (d) is greatly 
enlarged. This should be related to the special situation where the multishelled 
structure is lost and only two rows, i.e., an empty cylinder is present. After 
that, at a final stage of the shrinkage processes, one atomic chain remains.

\section{Conclusions}

MD computer simulation based on the well-established embedded-atom potential 
shows that finite gold wires of nanometre dimensions are often multishelled. 
Recently, similar gold nanostructures were formed in STM and observed by TEM 
\cite{Ohnishi}. 
The model of multishelled gold nanowires should be considered in 
explanation of the conductance measured in STM \cite{Ohnishi,Yanson}. 
Results of computer 
simulations enable fabrication of similar metallic nanowires which will be used 
in nanoelectronic and nanomechanical devices.

\clearpage

\begin{figure}
\caption{
Atomic positions for a nanowire with length $5.4$ nm, and a radius of $0.9$ 
nm.}
\label{fig1}
\end{figure}

\begin{figure}
\caption{
Top view of the MD box for a nanowire shown in Fig. 1. Here and in Fig. 
3 the whole thickness of the wire along its axis is shown after $7.1$ ns of 
simulation.}
\label{fig2}
\end{figure}

\begin{figure}
\caption{
Top view of the whole MD box for a nanowire of $689$ atoms, with length 
$2.6$ nm, and a radius of $1.2$ nm.}
\label{fig3}
\end{figure}

\begin{figure}
\caption{
Vertical slice through a multishelled gold nanowire: 
(a) for a nanowire from Ref. [9] ($588$ atoms, length $4$ nm, radius $0.9$ nm),
(b) for a nanowire shown here in Fig. 1 and Fig. 2. 
Vertical rows of gold atoms are the image of coaxial cylindrical shells. Similar 
structures were recently formed in STM and observed by an electron microscope 
[1].}
\label{fig4}
\end{figure}

\end{document}